\documentclass[usenatbib]{mnras}
\usepackage{times,graphicx,amssymb}

\title[Tidal parameters]{The tidal parameters of TRAPPIST-1 b and c}

\author[R. Brasser et al.]{R. Brasser$^1$, A. C. Barr$^2$ and V. Dobos$^{3,4,5}$\\
$^1$ Earth Life Science Institute, Tokyo Institute of Technology, Meguro, Tokyo 152-8551, Japan\\
$^2$ Planetary Science Institute, 1700 East Fort Lowell, Suite 106, Tucson, AZ, 85719, USA \\
$^3$ Konkoly Observatory, Research Centre for Astronomy and Earth Sciences, Hungarian Academy of Sciences, H-1121 
Konkoly Thege Mikl\'os \'ut 15-17, Budapest, Hungary \\
$^4$ Geodetic and Geophysical Institute, Research Centre for Astronomy and Earth Sciences, Hungarian Academy of 
Sciences, H-9400 Csatkai Endre u. 6-8, Sopron, Hungary\\
$^5$ MTA-ELTE Exoplanet Research Group, 9700, Szent Imre h. u. 112, Szombathely, Hungary}
\begin{document}
\maketitle
\begin{abstract}
The TRAPPIST-1 planetary system consists of seven planets within 0.05 au of each other, five of which are in a 
multi-resonant chain. {These resonances suggest the system formed via planet migration; subsequent tidal evolution 
has damped away most of the initial eccentricities. We used dynamical N-body simulations to estimate how long it takes 
for the multi-resonant configuration that arises during planet formation to break. From there we use secular theory to 
pose limits on the tidal parameters of planets b and c. We calibrate our results against multi-layered interior models 
constructed to fit the masses and radii of the planets, from which the tidal parameters are computed independently.} 
The dynamical simulations show that the planets typically go unstable 30 Myr after their formation. {Assuming 
synchronous rotation throughout} we compute $\frac{k_2}{Q} \gtrsim 2\times 10^{-4}$ for planet b and $\frac{k_2}{Q} 
\gtrsim 10^{-3}$ for planet c. Interior models yield $(0.075-0.37) \times 10^{-4}$ for TRAPPIST-1 b and $(0.4-2)\times 
10^{-4}$ for TRAPPIST-1 c. The agreement between the {dynamical and interior} models is not too strong, but is 
still useful to constrain the dynamical history of the system. We suggest that this two-pronged approach could be of 
further use in other multi-resonant systems if the planet's orbital and interior parameters are sufficiently well known.
\end{abstract}
\section{Introduction}

\begin{figure}
\resizebox{\hsize}{!}{\includegraphics[angle=0]{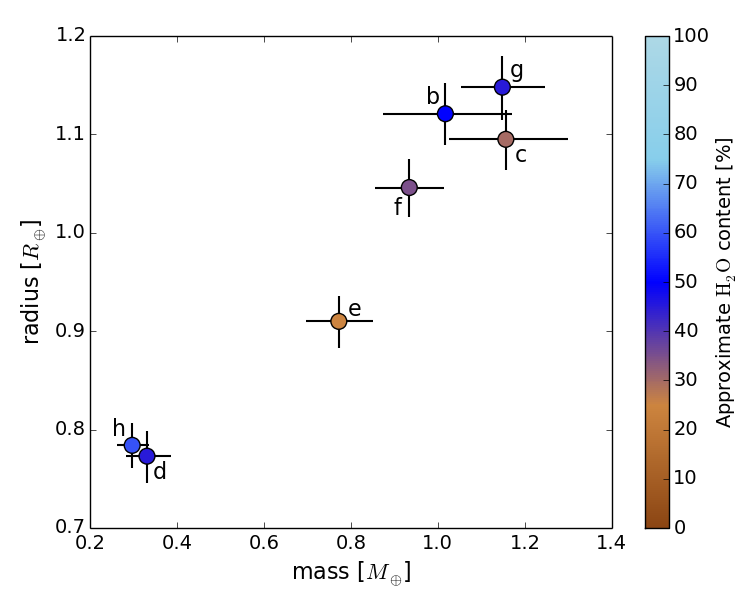}}
\caption{Mass-radius relationship of the TRAPPIST-1 planets from \citet{grimm18}. Colours represent estimated H$_2$O 
content (including liquid water, ice and high-pressure polymorphs) in volume fraction (percentage), based on the 
calculations for representative cases in the work of \citet{dobos19}.}
\label{fig:mrt1}
\end{figure}

The star TRAPPIST-1 is an ultracool M-dwarf that harbours seven roughly Earth-sized planets \citep{gillon17}. All of 
the planets orbit within 0.07 au of the star, and have orbital periods from 1.5 to 19 days \citep{gillon17, wang17, 
grimm18}. Their orbits are slightly eccentric, and the outer five planets are in a multi-resonant chain which probably 
maintains the orbital eccentricities over time \citep{unterborn18,luger17, grimm18}. All of the planets have a density 
intermediate between the densities of compressed water ice and the Earth's inner core \citep{gillon17, wang17, 
unterborn18, grimm18, barr18}, implying solid planets composed of rock, metal, and ice (see Figure~\ref{fig:mrt1}). 
Despite their proximity to the TRAPPIST-1 star, the current stellar flux on each planet is modest, suggesting effective 
black-body surface temperatures ranging from 400 K to 167 K \citep{wang17}. {Basic orbital and physical quantities for 
these planets are listed in Table~\ref{table:masses}.}

The planets' eccentric orbits and short orbital periods raise the possibility that tidal dissipation may be a 
significant heat source in their interiors \citep{luger17, barr18}, both now and in the past. \citet{barr18} showed 
that the planets' proximity to the central star and their eccentric orbits leads to interior geodynamics similar to 
that expected in the tidally heated satellites of the outer planets, specifically the inner Galilean satellites of 
Jupiter \citep{khurana2011}. Several of the TRAPPIST-1 planets could have partially molten rock mantles arising from a 
balance between heat generation by tides and heat transport by solid-state convection.  These conclusions remain valid 
despite recent updates to the masses and radii of the TRAPPIST-1 planets \citep{dobos19}.

Although the TRAPPIST-1 star is thought to be approximately 8 Gyr old, albeit with a large uncertainty \citep{BM17}, 
initial $N$-body simulations of the system's evolution have shown that the planets' orbits may only be stable in their 
present configuration for $\sim$ 0.5 Myr \citep{gillon17}. Torques between the planets and the TRAPPIST-1 star can 
cause changes in the semi-major axes of the planets, and eccentricity damping by tidal dissipation can cause the orbits 
to circularize. Eccentricity damping is known to enhance the system's stability, with the damping rate controlled by 
the planets' internal rigidity and viscosity through the value of the $k_2$ Love number, which describes how the 
planet's gravitational potential changes in response to tidal deformation, and the tidal quality factor, $Q$, which is 
a measure of how many orbital periods are needed to damp the tidal energy \citep{murray99}. The tidal torque 
additionally depends on the principal moment of inertia coefficient, $C=I/M_\mathrm{pl}R_\mathrm{pl}^2$, where $I$ is 
the moment of inertia around the planets' principal rotation axis, $M_\mathrm{pl}$ is the planet's mass, and 
$R_\mathrm{pl}$ is its radius \citep{MignardI}. In the orbital stability studies of \citet{gillon17} and 
\citet{luger17}, terrestrial and lunar values were used for the tidal parameters \citep{Lambeck80,neron97, bolmont15} 
for all of the TRAPPIST-1 planets. In reality the tidal quantities depend sensitively on the planets' interior 
structures and thermal states \citep[e.g.,][]{peale78, Lambeck80, segatz88}, {which have been shown to be 
substantially different from that of the Earth and Moon \citep{barr18}.}

Here, we independently calculate two estimates for the tidal parameters of the innermost two TRAPPIST-1 planets using a 
combination of N-body dynamics and the simple four-component compositional model from \citet{dobos19}. We provide 
estimates for the tidal quality 
factor $Q$ and $\frac{k_2}{Q}$ from both models that are consistent with the interior geodynamics, and discuss the 
agreement 
between the two approaches. 

\begin{table}
	\centering
	\caption{Semi-major axes ($a$), eccentricities ($e$), masses ($M_\mathrm{pl}$, scaled by Earth's mass, 
$M_{\oplus}=5.98\times10^{24}$ kg), and radii ($R_\mathrm{pl}$, scaled by Earth's radius, $R_{\oplus}=6371.8$ km) for 
the seven TRAPPIST-1 planets from \citet{grimm18} and \citet{delrez18}.}
	\label{table:masses}
	\begin{tabular}{lcccr} 
		\hline
		Planet & $a$ (AU) & $e$ & $M_\mathrm{pl}$ ($M_{\oplus}$) & $R_\mathrm{pl}$ ($R_{\oplus}$)\\
		\hline
        b & 0.01154775 & 0.00622& $1.017_{-0.143}^{+0.154}$ & $1.121_{-0.032}^{+0.031}$\\
        c & 0.01581512 & 0.00654 & $1.156_{-0.131}^{+0.142}$ & $1.095_{-0.031}^{+0.030}$ \\
        d & 0.02228038 & 0.00837 & $0.297_{-0.035}^{+0.039}$ & $0.784_{-0.023}^{+0.023}$ \\
        e & 0.02928285 & 0.00510 & $0.772_{-0.075}^{+0.079}$ & $0.910_{-0.027}^{+0.026}$ \\
        f & 0.03853361 & 0.01007 & $0.934_{-0.078}^{+0.080}$ & $1.046_{-0.030}^{+0.029}$ \\
        g & 0.04687692 & 0.00208 & $1.148_{-0.095}^{+0.098}$ & $1.148_{-0.033}^{+0.032}$ \\
        h & 0.06193488 & 0.00567 & $0.331_{-0.049}^{+0.056}$ & $0.773_{-0.027}^{+0.026}$ \\
		\hline
	\end{tabular}
\end{table}

\section{Current dynamical status of the TRAPPIST-1 system} \label{currentdyn}
\begin{figure*}
\resizebox{\hsize}{!}{\includegraphics[angle=0]{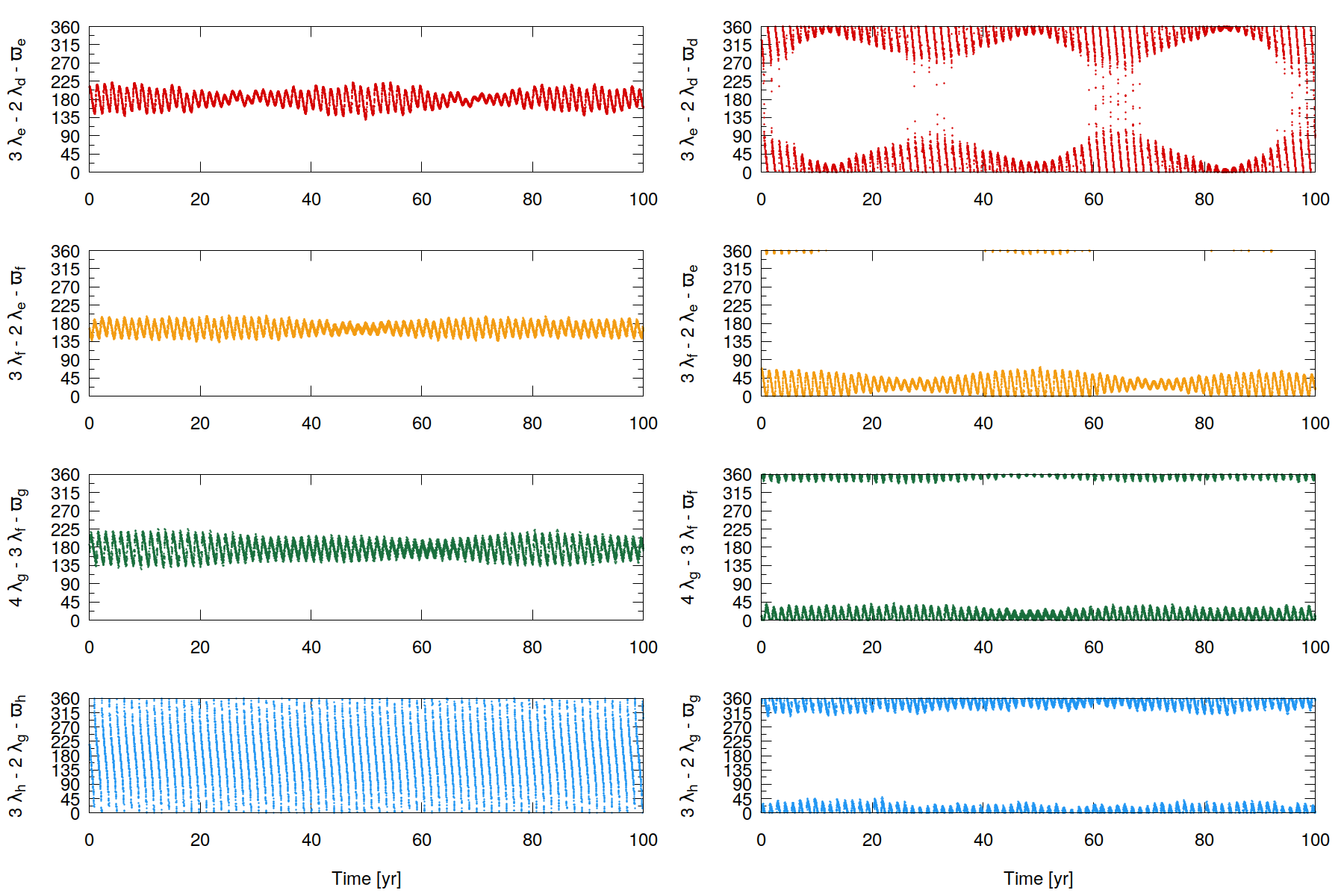}}
\caption{Evolution of the two-body resonant angles. With one exception in the bottom left panel showing the resonance 
angle $\phi_{\rm gh2}$, all the rest librate with a period $\sim$ 1.4 years.}
\label{fig:2bod}
\end{figure*}

The TRAPPIST-1 system is one of the few known systems in which most of the planets are in an orbital mean-motion 
resonance with each other. It is characterised by at least four two-body mean-motion resonances as well as up to five 
three-body resonances \citep{luger17}. We test whether these resonances still can be seen in the latest observed 
physical and orbital parameters (see Table~\ref{table:masses} and Figure~\ref{fig:mrt1}) from \citet{grimm18}. We 
simulate the system with nominal parameters for 100 years and 100 kyrs respectively with the SWIFT MVS software package 
\citep{LD94}. We do not investigate stability of the nominal system on longer timescales. The effect of general 
relativity is included for the periastron precession \citep{NW86}. The time step is set to 0.03625 days. We set the 
mutual inclinations ($i$) and longitudes of the ascending nodes ($\Omega$) to 0 for all planets. We find that planets d 
to g are trapped in sequential two-body mean-motion resonances, and a resonance also involves planet h since their 
resonant angles ($\phi$) librate. These resonances are
\begin{eqnarray}
 \phi_{\rm de1} &=& 3\lambda_e-2\lambda_d -\varpi_d, \nonumber \\
 \phi_{\rm de2} &=& 3\lambda_e-2\lambda_d -\varpi_e, \nonumber \\
 \phi_{\rm ef1} &=& 3\lambda_f-2\lambda_e -\varpi_e, \nonumber \\
 \phi_{\rm ef2} &=& 3\lambda_f-2\lambda_e -\varpi_f, \nonumber \\
 \phi_{\rm fg1} &=& 4\lambda_g-3\lambda_f -\varpi_f, \nonumber \\
 \phi_{\rm fg2} &=& 4\lambda_g-3\lambda_f -\varpi_g, \nonumber \\
 \phi_{\rm gh1} &=& 3\lambda_h-2\lambda_g -\varpi_g, \nonumber
\end{eqnarray}
\noindent where $\lambda = M + \varpi$ is the mean longitude, with $M$ being the mean anomaly, $\varpi = \omega + 
\Omega$ is the longitude of periastron and $\omega$ is the argument of periastron. The last remaining resonance angle, 
$\phi_{\rm gh2} = 3\lambda_h-2\lambda_g-\varpi_h$, does not librate. Figure~\ref{fig:2bod} shows the evolution of the 
resonant angles with time for 100~yr, indicating libration for all but one of the resonant angles. The libration period 
of the resonant angles is approximately 1.4 years.

\begin{figure*}
\resizebox{\hsize}{!}{\includegraphics[angle=0]{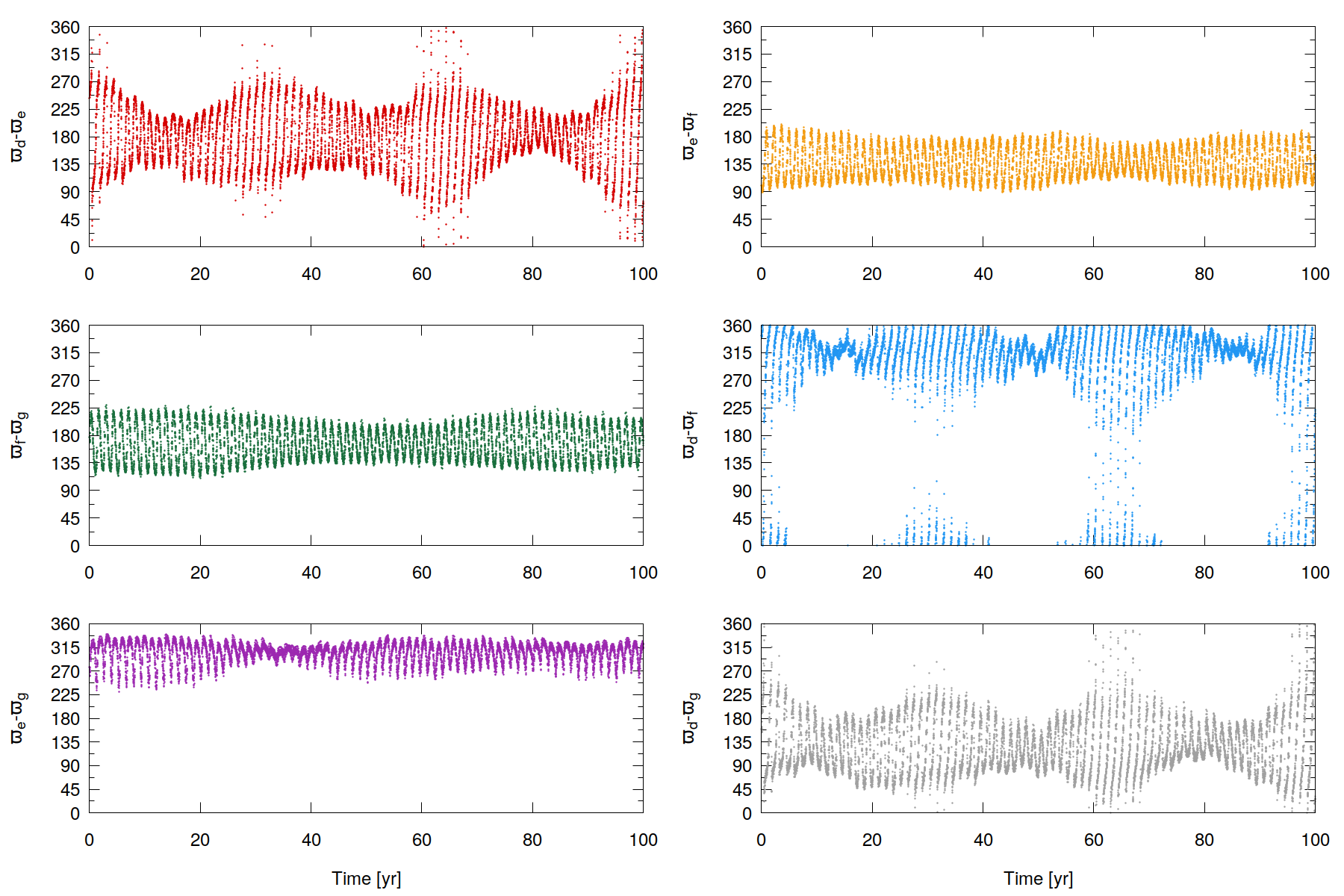}}
\caption{Apsidal anti-alignments for each planet pairs from planets d to g.}
\label{fig:dvarpi}
\end{figure*}

The resonant angles can be manipulated to compute $\Delta \varpi = \varpi_i-\varpi_j$ between any two planets. The 
results are displayed in Figure~\ref{fig:dvarpi}. Generally each consecutive pair of planets is approximately 
anti-aligned ($\Delta \varpi \sim 180^\circ$) but other non-sequential pairs are librating around values approximately 
60$^\circ$ away from 0. Alignment with planet d is sometimes broken into circulation because of secular forcing from 
planets b and c. Secular perturbations generally make the longitudes of periastrion progress with time ($\dot{\varpi} > 
0$) but in a mean motion resonance generally the periastra regress ($\dot{\varpi}<0$) \citep{MD99}. For the resonant 
planets the regression period is also equal to the libration period of the resonant angle i.e. 1.4 years, so that the 
regression is about $\dot{\varpi}= -951\,000$ ''/yr, or approximately 4.61 rad/yr. This is between 0.8\% to 3\% of the 
orbital motion of planets d to g. Planets b and c are not tied to any of the resonances and their longitudes of 
periastron progress with a period of about 40 years. For planet h the precession is 350 years. These precession periods 
correspond to frequencies of 30\,000 ''/yr and 3700 ''/yr respectively.

\begin{figure*}
\resizebox{\hsize}{!}{\includegraphics[angle=0]{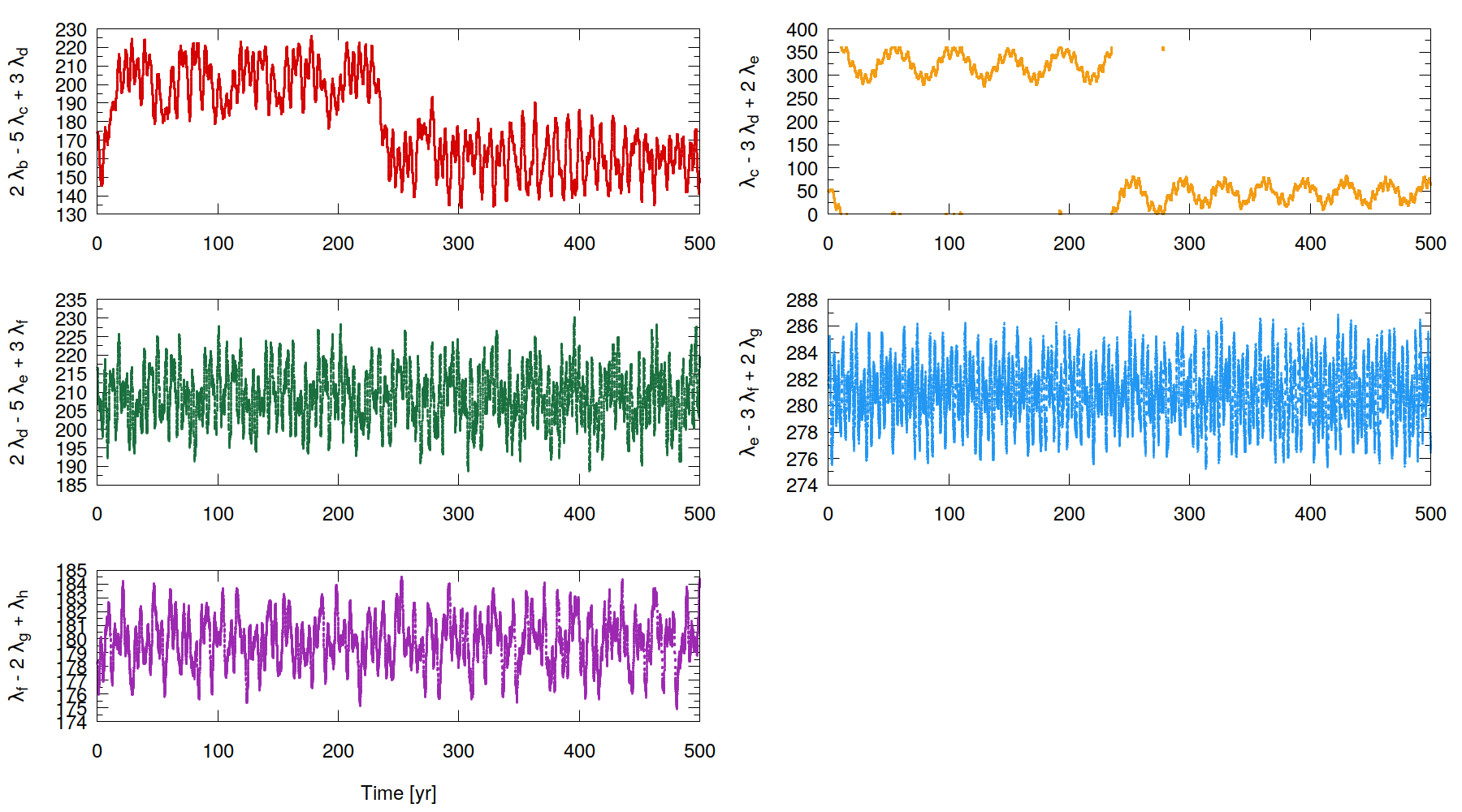}}
\caption{Three-body resonance angles on a short time-scale (500 years).}
\label{fig:3bods}
\end{figure*}

\begin{figure*}
\resizebox{\hsize}{!}{\includegraphics[angle=0]{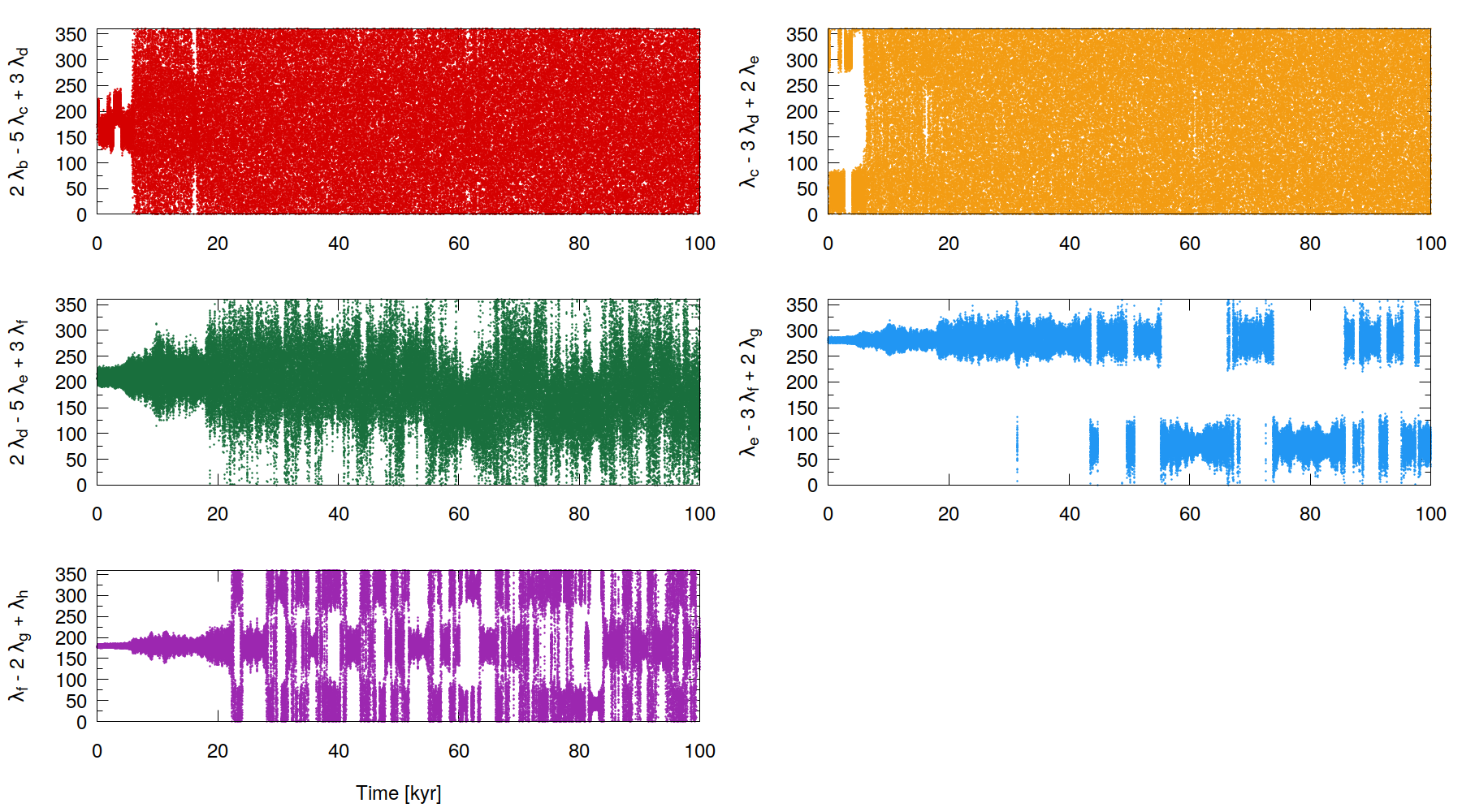}}
\caption{Three-body resonance angles on a long time-scale (100 000 years).}
\label{fig:3bodl}
\end{figure*}

The planets are also caught in several three-body resonances \citep{luger17}. These are
\begin{eqnarray}
\theta_{\rm bcd} &=& 2\lambda_b - 5\lambda_c + 3\lambda_d, \nonumber \\
\theta_{\rm cde} &=&  \lambda_c - 3\lambda_d + 2\lambda_e, \nonumber \\
\theta_{\rm def} &=& 2\lambda_d - 5\lambda_e + 3\lambda_f, \nonumber \\
\theta_{\rm efg} &=&  \lambda_e - 3\lambda_f + 2\lambda_g, \nonumber \\
\theta_{\rm fgh} &=&  \lambda_f - 2\lambda_g + \lambda_h.
\end{eqnarray}
\noindent and are the result of combining the angles of two consecutive two-body resonances. The short-term variation 
of these angles is plotted in Figure~\ref{fig:3bods}. On longer timescales, however, all of the three-body resonances 
break (see Figure~\ref{fig:3bodl}). We find that this typically happens after a few tens of thousands of years, much 
earlier than found by \cite{grimm18}. We do not know what methodology and initial conditions they employed to keep the 
three-body angles librate for more than a million years. In contrast, the two-body resonant angles appear stable for at 
least 100 kyr.

\begin{figure*}
\resizebox{\hsize}{!}{\includegraphics[angle=0]{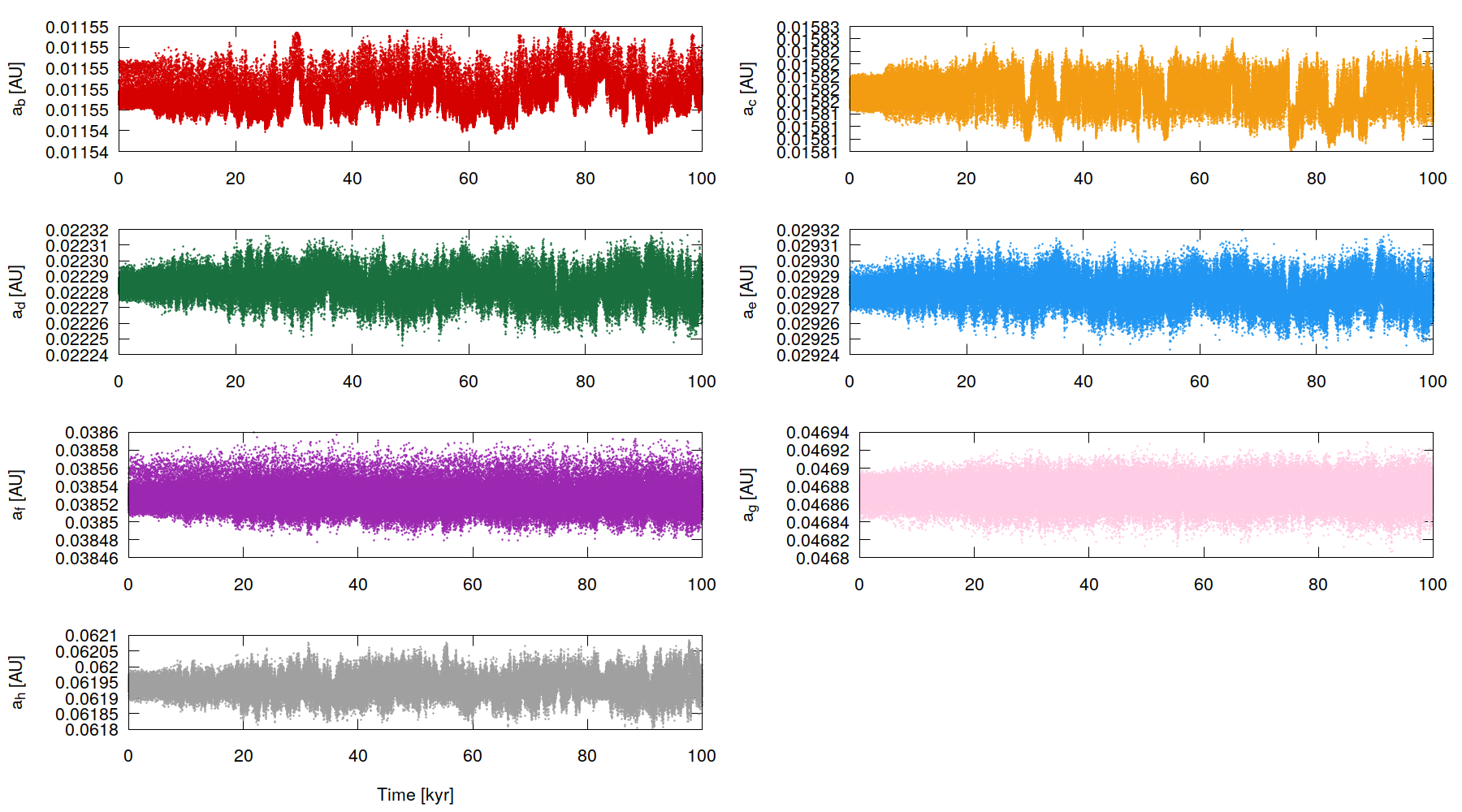}}
\caption{Semi-major axes of the planets' orbit on a long time-scale (100 000 years).}
\label{fig:along}
\end{figure*}

When the system leaves the three-body resonances, the semi-major axes of the planets exhibit irregular jumps 
(Figure~\ref{fig:along}). Unlike the planets in the solar system, for which the semi-major axes are constant, the 
TRAPPIST-1 system is caught in multiple resonances which affect the semi-major axes. Since none of the major planets in 
the solar system are resonant, the system is governed by secular interaction rather than resonant interaction, and the 
former preserves the semi-major axes \citep{MD99}. The irregular jumps in the semi-major axes of the TRAPPIST-1 planets 
are attributed to three-body resonance overlap \citep{Q11}, and the system is therefore chaotic. A very rough estimate 
of the longevity of the system against chaotic diffusion in semi-major axes is given by the time it takes for the 
semi-major axes to wander a distance approximately equal to the interplanetary spacing. Two estimates of this timing 
are given by \citep{Q11,QF14}
\begin{eqnarray}
\tau_{\rm l1} &\sim& 0.125\mu^{-3}\delta^6\vert \ln \delta \vert^{-3}, \nonumber \\
\tau_{\rm l2} &\sim& \mu^{-3}\delta^{15/2},
\end{eqnarray}
\noindent where $\mu = m_p/M_*$ and $\delta = \alpha_{\rm jk}^{-1}-1$ is the interplanetary spacing, with $\alpha_{\rm 
jk} = {\rm min}(a_{\rm j}/a_{\rm k},a_{\rm k}/a_{\rm j})$ depending on the indices $j$ and $k$. For TRAPPIST-1 
typically $\mu \sim 2 \times 10^{-5}$ and $\delta \sim 0.3$ so that $\tau_{\rm l1} = 6.7$~Gyr and $\tau_{\rm l2} = 
15$~Gyr. A stability time of 10 Gyr would require $\delta > 0.27$ \citep{FQ07}, which is satisfied for all pairs except 
f and g. Given that the age of the system is about 8 Gyr \citep{BM17}, in the absence of tidal relaxation, 
the system could be close to instability, although the low angular momentum deficit (AMD) of the system would prevent a 
global instability \citep{laskar1997}.

\section{The possible formation of the TRAPPIST-1 system}
\cite{Ormel17} have performed the most in-depth study of the formation of the TRAPPIST-1 system based on pebble 
accretion. They conclude that a mixture of said accretion and type 1 migration will place the planets in a 
multi-resonant configuration. The lack of resonances among the inner three planets is readily explained by the inner 
planets having been nudged outwards by the edge of the protoplanetary disc due to magnetic rebound \citep{Liu17}. While 
this demonstrates several arguments in favour of forming the TRAPPIST-1 system with pebble accretion, the study by 
\cite{Ormel17} just assumes that the planets migrate into resonances and forgoes performing a more in-depth study of 
the dynamics of this resonant trapping and its consequences. While a very detailed analysis of this effect is a study 
in itself, here we build a plausible argument for what the orbital structure of the TRAPPIST-1 system could have been 
before and shortly after the dispersal of the protoplanetary disc.

For simplicity we assume steady-state accretion of the disc gas onto the star. The gas accretion rate is related to the 
gas surface density and scale height of the disc via
\begin{equation}
\dot{M}_*= 3\pi \alpha_{\rm acc} \Sigma H^2 \Omega_{\rm K},
\label{eq:dotmstar}
\end{equation}
\noindent where $\Sigma$ is the gas surface density, $H$ is the disc scale height and $\Omega_{\rm K}$ is the Kepler 
frequency. The viscosity $\nu=\alpha_{\rm acc} c_s^2\Omega_{\rm K}$ \citep{SS73}, where $\alpha_{\rm acc}$  is an 
`effective' parameter for global angular momentum transfer of the disc, which is assumed to be constant. The disc scale 
height is related to the temperature via $H=c_s/\Omega_{\rm K}$ where $c_{\rm s} ^2=(k_BT/\mu m_p)$ is the isothermal 
sound speed, $k_B$ is the Boltzmann constant, $m_p$ is the proton mass, and $\mu=2.3$ is the mean atomic mass of the 
gas.

For simplicity we make use of the disc model from \citet{ida16}, which is based on the works of \citet{GL07} and 
\citet{Oka11}, but the analysis below can also be applied to more complex disc models. The disc is assumed to be in a 
steady state and the temperature and surface density are power laws of the distance to the star. The best fit for the 
temperature profile is given by \citep{GL07,ida16}
\begin{equation}
 T = 150L_*^{2/7}M_*^{-1/7}\Bigl(\frac{r}{1\,{\rm AU}}\Bigr)^{-3/7}\; {\rm K}.
\label{eq:T_visirr}
\end{equation}
\noindent The unit of the stellar mass and stellar luminosity are the current solar values. The disc is assumed to be 
heated by the stellar flux, which generally applies to the outer portions of the disc. From the above temperature 
relation the reduced scale height becomes
\begin{equation}
h=0.025 L_*^{1/7}M_*^{-4/7}\Bigl(\frac{r}{1\,{\rm AU}}\Bigr)^{2/7}.
\label{eq:h_visirr}
\end{equation}
\noindent The surface density then follows from the steady state accretion and we have $p=-d\ln \Sigma/d \ln r = 
15/14$. The nominal luminosity of pre-main sequence M-dwarf stars is $L_* \sim M_*^2$ and thus in the first 1-10~Myr 
after the birth of TRAPPIST-1 the luminosity is $L_* \sim 0.01\,L_\odot$ \citep{Baraffe15}. A value of $\alpha_{\rm 
acc}=10^{-3}$ is consistent with the minimum-mass solar nebula (MMSN) when the stellar accretion rate is 
$\dot{M}=10^{-8}$ $M_\odot$ yr$^{-1}$ \citep{ida16}, which corresponds to a stellar age of about 1~Myr \citep{H98}. 
Furthermore, the stellar accretion rate $\dot{M}_* \propto M_*^2$ \citep{Manara15}. For the young TRAPPIST-1 we then 
have that at 1~Myr $\dot{M}\sim 10^{-10}$ $M_\odot$ yr$^{-1}$ and $h \sim 0.023$ at 0.06~AU, and with these parameters 
our disc model is generally valid when $r \gtrsim 0.06$~AU. For comparison, \cite{Ormel17} adopted a constant reduced 
scale height of 0.03, which with our prescription occurs at about 0.15~AU. \cite{unterborn18} adopted a nearly 
identical 
disc prescription to ours.

It is likely that the TRAPPIST-1 planetary system migrated inwards to its current location \citep{unterborn18}. If this 
system formed through pebble accretion \citep{Ormel17}, then the migration was likely stalled by the innermost planet 
reaching the magnetic truncation of the disc \citep{Liu17} and the other planets all migrated into subsequent $j + 1:j$ 
mean-motion resonances. Resonance trapping occurs if the migration rate is slower than some critical value \citep{P13}, 
i.e. when the rate of change of the mean motion $\dot{n} \leq \dot{n}_{\rm crit}$, where $n^2=G(M_*+m_p)/r^3$ is the 
orbital frequency. This can be converted to a migration timescale $\tau_{c} = n/\dot{n}$
\begin{equation}
 \tau_c \geq \frac{26}{w_p}\frac{P}{10\,{\rm d}}\Bigl(\frac{M_\oplus}{m_p}\Bigr)^{4/3}\Bigl(\frac{M_*}
 {M_\odot}\Bigr)^{4/3}\, 
{\rm kyr},
\end{equation}
\noindent where $w_p$ is a numerical factor that depends on the resonance (1.33 for the 2:1 and 3.38 for the 3:2). If 
the planets are formed through pebble accretion \citep{Ormel17} then they are suggested to stop accreting solids when 
they reach the pebble isolation mass \citep{Lam14} $m_{\rm iso}=0.5h^2 M_* \sim 1\,M_\oplus$. This is comparable to the 
thermal mass at which the planet opens a gap in the disc, $m_{\rm gap} = 6m_{\rm iso} \sim 5 M_\oplus$. Planets  above 
this mass will execute so-called type II migration \citep{LP86}, while lower-mass planets execute so-called type I 
migration \citep{tanaka02}. Since all planets are closer together than the 2:1 resonance, but usually not closer than 
the 3:2, this sets both an upper and a lower limit on the migration speed which is inconsistent with type II.

To lowest order in eccentricity, the type I migration timescale of the planets is given by
\begin{equation}
\tau_a^{-1} = 2\tau_m^{-1} + 2e^2\tau_e^{-1} ,
\end{equation}
\noindent where the eccentricity damping time scale is \citep{cn08}
\begin{equation}
\tau_{e} = 1.282t_{\rm wav}(1-0.14\hat{e}^2+0.06\hat{e}^3),
\end{equation}
\noindent where $\hat{e} = e/h$. We further have $\tau_m = -L/\dot{L}$ which is given by
\begin{equation}
\tau_m = -\frac{t_{\rm wav}}{\Gamma h^2}P_e,
\end{equation}
\noindent where $\Gamma=\dot{L}$ is the total torque (usually negative) and $\tau_m<0$ for outward migration. Here 
$P_e$ is a function of eccentricity that takes care of supersonic corrections \citep{cn08}
\begin{equation}
P_e = \frac{1+(0.444\hat{e})^{1/2}+(0.352\hat{e})^6}{1-(0.495\hat{e})^4}.
\end{equation}
\noindent The wave timescale is given by \citep{TW04}
\begin{equation}
t_{\rm wav} = \Bigl(\frac{M_*}{m}\Bigr)\Bigl(\frac{M_*}{\Sigma r^2}\Bigr)h^4 n^{-1}.
\end{equation}
\noindent The total torque is the sum of the Lindblad and corotation torques. Accounting for the effect of saturation, 
in the subsonic case we have approximately $\Gamma = 5.51q -4.94$ \citep{P11} where $q=-\frac{d \ln T}{d \ln r}$, whose 
nominal value is 3/7. At 0.1~AU for an 1~$M_\oplus$ planet $t_{\rm wav} \sim 40$~yr and the typical migration time is 
$\tau_a \sim 12$~kyr. The critical migration time $\tau_c = 1150$~yr, so the planets are expected to get trapped into 
resonances (since $\tau_a > \tau_c$). In fact, it is expected that they should all be in the 2:1 resonance and the 
reason they passed over this resonance is because that configuration was probably overstable \citep{GS14}. Once capture 
occurs, 
the inward migration of the planets is balanced by the resonances and any planet pair will be caught in a resonance 
wherein their equilibrium eccentricity is \citep{GS14}
\begin{equation}
 e_{\rm eq}^2 = \frac{\tau_e}{(3j+1.282)\tau_a}.
\end{equation}
\noindent This leads to
\begin{equation}
 \frac{\tau_e}{\tau_a} = 2.564\vert\Gamma\vert h^2.
\end{equation}
\noindent With the nominal parameters, $e_{\rm eq} \approx 1.3h$. Thus we expect the TRAPPIST-1 system to have migrated 
into a multi-resonant $j+1:j$ chain, with planets b and c possibly having been dislodged by a magnetic rebound 
\citep{Ormel17}, where the eccentricities of all planets were comparable to the reduced scale height i.e. about 0.03.

\section{Numerical simulations of long-term stability}
The current eccentricities are about a factor of 3-5 lower than what is predicted when the planets were initially 
captured in resonance. Thus we theorise that the current configuration could have only been reached after tidal damping 
of the initial eccentricities. However, the key to stability is the timescale of eccentricity damping. From numerical 
experiments lasting 100~Myrs, \cite{I17} found that in excess of 75\% of multi-resonant low-mass planetary systems go 
unstable in the absence of tidal damping; based on observational data from exoplanetary systems they further argue that 
as many as 95\% multi-resonant systems must go unstable \citep[cf.][]{GS14}. This begs the question as to whether the 
primordial TRAPPIST-1 system, wherein the planets are all expected to have eccentricities $e \approx h$, was unstable 
as well, and if so, on what timescale. To test this hypothesis we perform simulations of the TRAPPIST-1 system for 100 
Myrs wherein we systematically increase the initial angular momentum deficit (AMD) of the system. We choose this 
quantity rather than the individual eccentricities of the planets because the former is a measure of the dynamical 
excitation of the whole system rather than that of an individual planet. The normalised AMD is given by
\begin{equation}
{\rm AMD} = \frac{\sum_k \mu_k \sqrt{a}_k (1-\sqrt{1-e_k^2})}{\sum_k \mu_k \sqrt{a}_k}.
\end{equation}
\noindent Its value for the current orbital parameters is approximately 2.6$\times$10$^{-5}$. We increase the initial 
AMD from 4 times to 21 times the current value, with the total AMD randomly partitioned amongst the planets; the upper 
value corresponds to all planets having an eccentricity of $e \sim h$. In these simulations we keep initial phases the 
same as those published in \cite{grimm18} to preserve the resonant structure. This allows us to investigate the effect 
of only the initial AMD on the instability timescale. The simulations are once again run with SWIFT MVS with the same 
parameters as in Section~\ref{currentdyn}, but the total simulation time is increased to 100 Myrs. A system is defined 
to be unstable when we record an encounter between a pair of planets.

\begin{figure*}
\resizebox{\hsize}{!}{\includegraphics[angle=0]{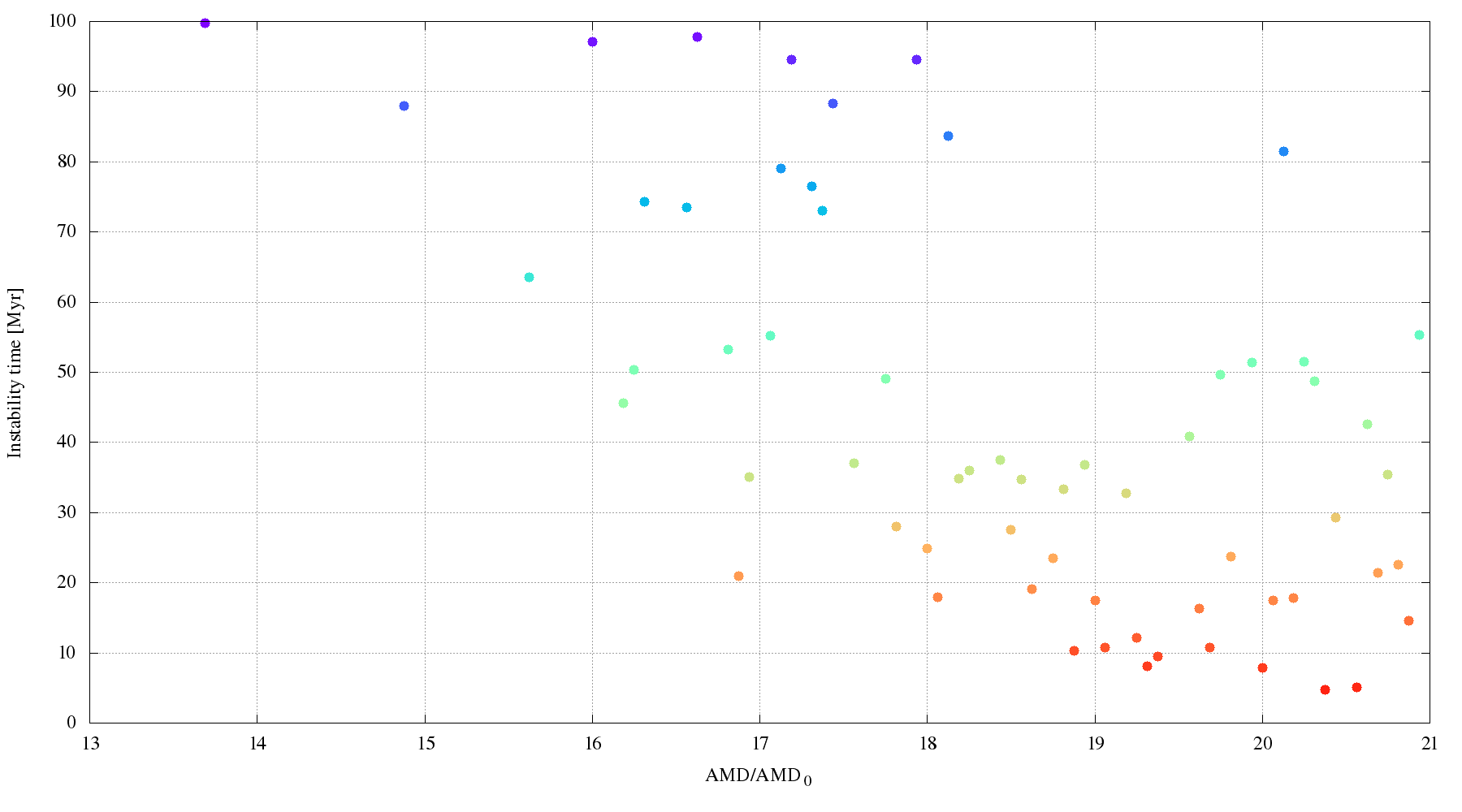}}
\caption{Instability time as a function of original AMD. The colours are there to aid visualisation.}
\label{fig:itime}
\end{figure*}

The outcome of these simulations are summarised in Figure~\ref{fig:itime}, which shows the timescale to reach 
instability versus the initial AMD normalised by the current value. We find that half of all unstable cases have 
AMD/AMD$_0\lesssim$ 19, and that the median time for the system to go unstable is 27.5 Myrs. There is large scatter in 
the plot but there is also a clearly decreasing trend for the onset of instability with increasing AMD. Systems with 
AMD/AMD$_0\leq$ 13 are generally stable for the duration of the simulation. Therefore, in absence of tidal dissipation, 
we expect the system to be unstable on a timescale of about 30 Myrs if all the planets originally had eccentricities 
comparable to the disc scale height. This sets an upper limit on the eccentricity damping timescale for the planets, 
and therefore on their interior structure and composition.

\section{Tidal evolution and eccentricity damping timescales}
To lowest order in eccentricity, the secular interaction between the planets follows the Laplace-Lagrange theory. This 
theory treats the planets as a set of coupled oscillators and can thus be reduced to an eigenproblem \citep{MD99}. Here 
we follow the elegant approach of \cite{BL11} and \cite{Lovis11} wherein tidal evolution and precession of the periapse 
caused by tides and general relativity are considered too. Unfortunately, the Laplace-Lagrange theory does not work for 
resonant planets, because the resonant interaction increases the precession rate close to resonance and reverses its 
direction in resonance \citep{MD99}. In Section~\ref{currentdyn} we showed that planets d to g are in multiple resonant 
chains, and that planets g and h are also in a resonance wherein one of the resonant angles librates. We deduced from 
our numerical simulations that the influence of the resonances causes the periapses of planets d to g to regress with 
peroids of about 1.4 yrs. Secular coupling is the strongest between planets whose precession frequencies are close 
together and in the same direction \citep[e.g.][]{MD99}. As a result, secular coupling between the resonant quadruplet 
d to g is very strong and they form one secular group. However, these middle planets are only weakly coupled to planets 
b, c and h, which form the second secular group, because of the large disparity (and opposite direction) of the motion 
of the periapses of these two planet groups. 

\begin{figure*}
\resizebox{\hsize}{!}{\includegraphics[angle=0]{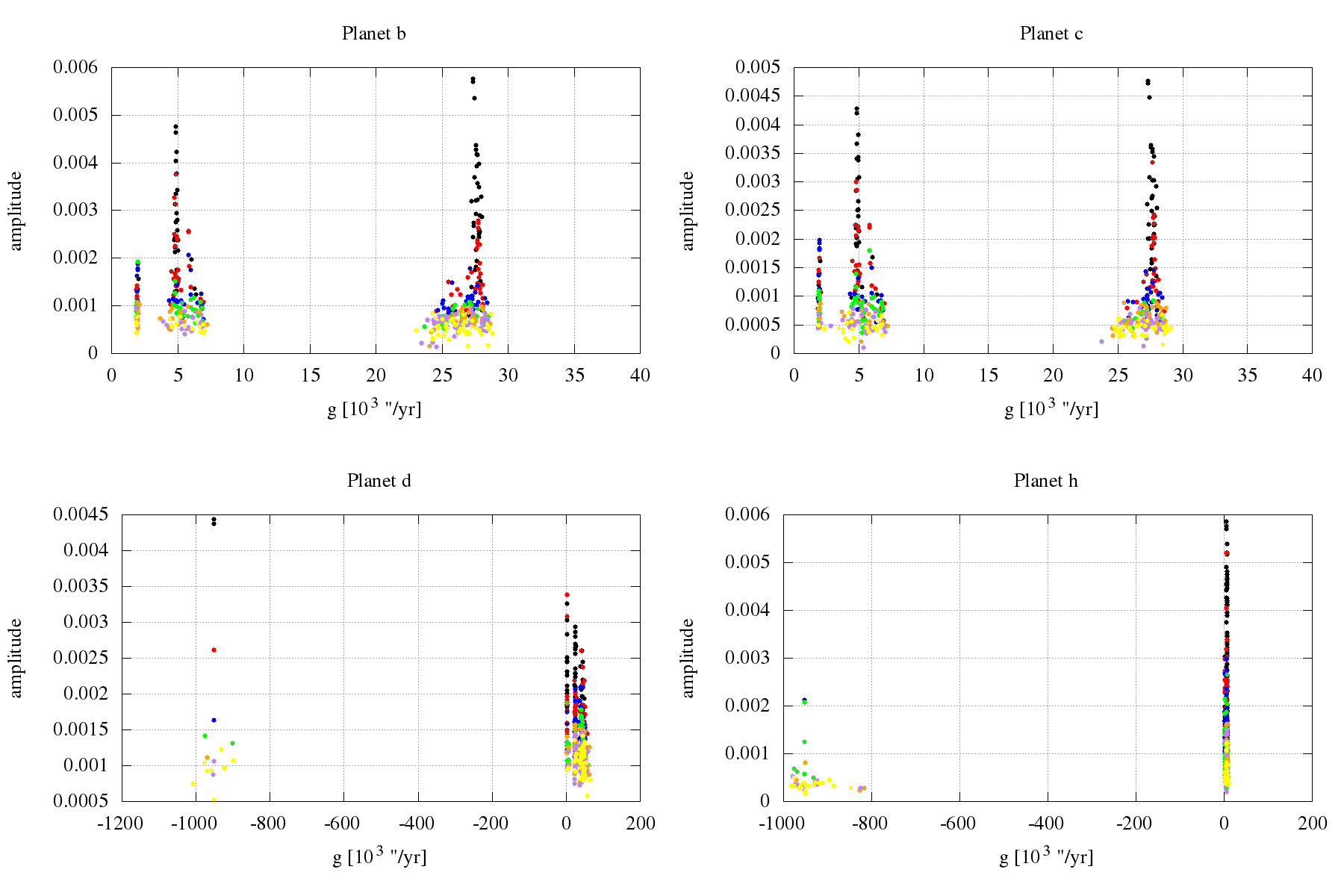}}
\caption{Fourier output of the eccentricity vectors of planets b, c, d and h for 64 intervals of 4096 years each. Only 
the 7 strongest modes were kept (depicted in black, red, blue, green, orange, indigo and cyan, respectively).}
\label{fig:fft}
\end{figure*}
To emphasise the validity of this claim Figure~\ref{fig:fft} shows the Fourier decomposition of the eccentricity 
vectors of planets b, c, d and h. Since the motion appears to be chaotic, we chose to show the decomposition over 64 
intervals of 4096 years each, for each planet, keeping only the 7 leading terms in the Fourier spectrum. Prominent 
frequencies are at 2000 ''/yr, 4900 ''/yr, 5800 ''/yr, 6500 ''/yr and 29500 ''/yr. It is clear that for planet d the 
leading terms are associated with the resonance, but that other terms also play a role; the resonant term also shows up 
in the spectrum of planet h. For the innermost two planets, however, the resonant terms do not appear. As such, we 
resort to approximate the secular interaction for the whole system with the Laplace-Lagrange theory with the caveat 
that 
only the eigenfrequencies associated with planets b and c, and possibly h, are useful. The effect of the three-body 
resonances on the precession frequencies is similar to the two-body resonances \citep{QF14}.

{Before we proceed a clarification is in order. In what follows we assume that the eccentricity evolution primarily 
occurs when the planets are in synchronous rotation. This is not entirely correct, because the tidal evolution on the 
planets will reduce both their eccentricities and their spin rates, and the planets may temporarily be trapped in 
spin-orbit resonances while their eccentricities are decreasing \citep{Corr14}. However, the initial eccentricities 
after formation are expected to be of order 0.03, which are low enough that only temporary capture in the 3:2 
spin-orbit resonance becomes feasible \citep{Corr14}. An in-depth study of how the system evolves tidally with the 
planets locked in this spin-orbit configuration is beyond the scope of this paper.}

Defining $z = e\exp(\imath \varpi)$, where $\imath$ is the imaginary unit and $\varpi=\Omega+\omega$ is the longitude 
of periapse, without tides the equation of motion for the eccentricity and periapse of planet $j$ is
\begin{equation}
 \dot{z}_j = \sum_{k=1}^N \imath A_{jk}z_k,
 \label{eq:znt}
\end{equation}
\noindent where the matrix ${\bf A}$ contains the mutual interaction terms, which only depend on the masses and 
separation \citep{MD99}, and the effect of general relativity \citep{BL11,Lovis11}. This system of equations has the 
solution 
\begin{equation}
z_j = \sum_{k=1}^N \bar{e}_{jk}\exp[\imath (g_kt + \delta_k)],
\label{eq:zsol}
\end{equation}
\noindent where the $g_k$ are the eccentricity eigenfrequencies and the $\bar{e}_{jk}$ are the eigenvectors and 
$\delta_k$ are initial phases.

To add in the tidal influence, we assume the eccentricities are damped according to $\dot{z}_j = z_j/\tau_t^{(j)}$ 
\citep{BL11}, where $\tau_t$ is the tidal damping timescale for each planet. {It can be demonstrated from equation 
(159) in \citet{be2019} that for synchronous planets the frequency dependencies of the quadrupole quality functions 
$k_{2j} /Q_j$ are model-independent. Consequently, for synchronous planets, the eccentricity damping 
timescales introduced by \citep{GS66} are also model-independent, and are given by}
\begin{equation}
 \tau_t^{(j)} = \frac{P_j}{21\pi}\frac{m_j}{M_*}\Bigl(\frac{R_j}{a_j}\Bigr)^{-5}\Bigl(\frac{k_{2j}}{Q_j}\Bigr)^{-1}
 \label{eq:tedamp}
\end{equation}
\noindent where $P_j$ is the planet's orbital period, $m_j$ is the mass, $k_{2j}$ is the second degree Love number, 
$Q_j$ is the tidal dissipation parameter, $R_j$ is the radius and $a_j$ is the semi-major axis of planet $j$. 
The effect of eccentricity damping by tidal friction within the planets can be modelled by adding an additional term to 
equation 
(\ref{eq:znt}) as
\begin{equation}
 \dot{z}_j = \sum_{k=1}^N (\imath A_{jk} + C_{jk})z_k,
 \label{eq:zt}
\end{equation}
\noindent where $C_{jj} = 1/\tau_t^{(j)}$ \citep{BL11}. The diagonal matrix ${\bf C}$ adds a complex component to the 
eigenfrequencies. The solution is the same as equation~(\ref{eq:zsol}), but eigenfrequencies $g_k$ are now complex and 
the eigenvectors decrease with time. Furthermore, the imaginary components of the eigenfrequencies are unequal, so that 
the decay timescale of several of the modes can be considerably longer than the shortest ones. Hence the system will 
eventually evolve to a state that has only a single secular eigenmode. The decay timescale of each eigenmode is 
\citep{BL11}
\begin{equation}
\tau_d^{(j)} = [{\rm Im}(g_j)]^{-1}.
\end{equation}
\noindent The eigenmode decay timescale $\tau_d^{(j)}$ can be far longer than the tidal damping timescale 
$\tau_t^{(j)}$ so that even the innermost planets' orbits can stay eccentric for a long time because of secular 
coupling 
with the more distant planets \citep{BL11,Lovis11}. Once the system only contains a single secular eigenmode, the rates 
of orbital precession are identical for all planets that are not in resonance. These orbits are either aligned or 
anti-aligned, depending on which particular eigenmode has survived. The orbital elements of the current system indicate 
that it has not yet reached that state. This is expected for planets d to g because they are in (three-body) 
resonances. 
The eccentricities of planets b, c and h are also inconsistent with zero, and of these only planet h is caught in a 
resonance.

\begin{table}
 \begin{tabular}{cccc}
  \hline
  planet & Eigenfrequency [''/yr] & $\tau_d^{(j)}$ [$\tau_d^{(b)}$] & $\tau_t^{(j)}$ [$\tau_t^{(b)}$]\\
  \hline
  b & 27300 ($g_1$)& 1.00 & 1.00 \\
  c & 4450 ($g_2$) & 1.58 & 5.62\\
  d & 3370 ($g_3$) & 6.82 & 82.7\\
  e & 8920 & 9.73 & 203\\
  f & 16500 & 15.0 & 521\\
  g & 15400 & 25.2 & 1060\\
  h & 1410 ($g_7$) & 30.9 & 13700\\
  \hline
 \end{tabular}
\caption{Eigenfrequencies and eigenmode damping timescales. All of these frequencies are about 10\%-30\% lower than 
those computed with Fourier analysis from N-body simulations, primarily caused by near-resonances between planets b and 
c, planets c and d, and the circulation of $\phi_{\rm gh2} = 3\lambda_h-2\lambda_g -\varpi_h.$}
\label{tab:evd}
\end{table}

The eigenfrequencies and associated eigenmodal damping times are listed in Table~\ref{tab:evd} as a multiple of the 
secular damping timescale in planet b ($\tau_d^{(b)}$). Here we assumed that the lowest value of $\tau_d^{(j)}$ 
corresponds to dissipation in planet b, the next shortest by dissipation in planet c, and so forth. Naturally some of 
the calculated frequencies make little sense because the resonant interaction is not taken into account, but for the 
frequencies corresponding to planets b, c and h this assumption is justified. The results in the table are based on an 
additional simplifying assumption that all the planets have the same composition {and therefore the same rheology}. 
This is demonstrably false \citep{unterborn18, barr18, dorn18}, but this simple example serves only as an indicator of 
the damping timescale disparity. {Assuming that all the planets' interior temperatures are just above the solidus, 
their viscosities are $\eta \sim 10^{14}$~Pa and rigidity $\mu \sim 10$~GPa \citep{barr18}. Following \citet{Corr14} we 
compute theoretical values of the tidal damping timescales $\tau_t^{(j)}$ and modal damping timescales $\tau_d^{(j)}$. 
The actual values are, for now, unimportant; at present we are interested in their relative values only.}

From this simple experiment it is clear that the mode that damps the slowest takes at least 30 times as long as the 
mode that damps the fastest. {This is in stark contrast to the ratio of the physical damping timescales in each of 
the planets, where the maximum ratio runs into the ten thousands.} Significantly increasing or decreasing the damping 
efficiency in planets c to h while keeping the parameters for planet b fixed does not substantially change the values 
reported in the table, implying that most of the damping of the {\it whole system} occurs in planet b (and c). {The 
disparate ratios of the values of $\tau_t^{(j)}$ compared with $\tau_d^{(j)}$ reinforce this claim.} Unless damping in 
planet b is much less efficient than in planet c, the rate of eccentricity dissipation in planet c is expected to be 
about a third of that in planet b if they share a similar composition and tidal parameters.

We argue that if we want to keep the system dynamically stable within 100 Myrs to avoid breaking the resonance 
\citep{I17}, and in light of the result given in Figure~\ref{fig:itime}, the damping timescale in planet b cannot 
greatly exceed 10~Myrs, because we need to damp most of the AMD within 30 to 100 Myrs. This sets a lower boundary on 
the tidal parameters of this planet. Generally $\tau_d^{(j)} \neq \tau_t^{(j)}$, but for planet b they are 
comparable within a factor of two for a given composition because the term $C_{11}$ dominates the secular motion 
of planet b in equation (\ref{eq:zt}). Assuming these timescales are equal, {assuming synchronous rotation and 
setting $\tau_t^{(b)}\lesssim 10$~Myr} implies that for planet b we have $\frac{k_2}{Q} \gtrsim 2 \times 10^{-4}$, 
which is lower than the measured values of Mars, the Moon and of the deep Earth. If we further assume that $\tau_d^{(c)} 
= \tau_t^{(c)}$, which is also correct within a factor of two {for the rheologies assumed here}, then {setting 
$\tau_t^{(c)}\lesssim 10$~Myr} for planet c we have $\frac{k_2}{Q} \gtrsim 1 \times 10^{-3}$, which is comparable to 
that in the Moon, Mars and deep Earth. In the next section we compare these values with those generated from interior 
models.

\section{Tidal parameters from interior models}
The response of a planet to tidal forcing is described by the three Love numbers, $h_2$ and $l_2$, which describe tidal 
deformation, and $k_2$, which describes changes to the planet's own gravitational potential due to the tidal 
deformation. For a viscoelastic body \citep{peltier74, greff05, henning09, wahr09, efroimsky12, barr18},
\begin{equation}
k_2^*=\frac{3/2}{1+\frac{19\mu^*}{2\bar{\rho}g R_\mathrm{pl}}}, \label{eq:k2}
\end{equation}
where $\bar{\rho}$ is the planet's bulk density, $g$ is its surface gravity, and the planet's rigidity is $\mu^*=M_1 + 
i M_2$, where the star superscript denotes complex quantities. {For simplicity we apply the Maxwell viscoelastic 
body model \citep{peltier74},  for which}
\begin{equation}
M_1=\frac{\mu \nu^2 \eta^2}{\mu^2 + \nu^2 \eta^2}, \label{eq:M1}
\end{equation}
and
\begin{equation}
M_2=\frac{\mu^2 \nu \eta}{\mu^2 + \nu^2 \eta^2}, \label{eq:M2}
\end{equation}
where $\mu$ is the rigidity, $\eta$ is viscosity, and $\nu=2n-2\Omega_{\rm rot}$ is the tidal forcing frequency, in 
this case, $\nu=2\pi/P$, where $P$ is the orbital period of each planet. {The Maxwell model does not adequately 
reproduce the tidal dissipation in cold planets such as Mars \citep{Bills2005}, but the innermost few TRAPPIST-1 planets 
are expected to have hot interiors \citep{barr18} so that the application of the Maxwell model may be warranted.} The 
overall behaviour of the planet is analogous to a damped, driven oscillator, being driven at frequency $\nu$. 
Dissipation is maximized when the forcing time scale, $T_\mathrm{f} =  2 \pi/\nu$ is equal to the {Maxwell time, 
which is} ratio between the viscosity and rigidity, i.e. $\tau_M=\eta/\mu$.

We mimic the effect of a multi-component, layered planet  using a single, approximate uniform viscosity and rigidity 
\citep{barr18, dobos19},
\begin{equation}
\eta(T) \approx \phi_\mathrm{iw}\eta_\mathrm{iw} + \phi_\mathrm{hpp} \eta_\mathrm{hpp} + \phi_\mathrm{r} 
\eta_\mathrm{r}.
\end{equation}
where $\eta$ represents the viscosity of the material. The indices refer to different material layers in the body: 
``iw'' for ice I and/or liquid water, ``hpp'' for high-pressure polymorphs, ``r'' for rock, and ``Fe'' for iron. We use 
a similar relationship to construct a single value of the shear modulus ($\mu$) that approximates the behaviour of the 
entire planet \citep{barr18, dobos19},
\begin{equation}
\mu(T) \approx \phi_\mathrm{iw}\mu_\mathrm{iw} + \phi_\mathrm{hpp} \mu_\mathrm{hpp} + \phi_\mathrm{r} 
\mu_\mathrm{r}.
\end{equation} 
We neglect the viscosity and rigidity of the planets' iron cores because tidal deformation of, and thus dissipation in, 
the cores will be negligible due to the restoring force imposed by the rock and ice mantle that lies atop the core in 
each body \citep{henning09}. 

The $\eta$ and $\mu$ for each of the constituent materials depend strongly on temperature. Governing parameters and 
equations describing the material behavior may be found in \citet{barr18} and are too cumbersome to repeat here in full 
detail. For the rocky component of the TRAPPIST-1 planets, we assume a compressed bridgmanite composition 
\citep{unterborn18}, consistent with the lower mantle of the Earth, with a density $\rho_{\rm r} = 5500$ kg/m$^3$ 
\citep{dziewonski81}. The viscosity and rigidity of rock change as a function of temperature, with both quantities 
decreasing sharply for temperatures above the solidus ($T_{\rm s}=1600$ K).  Another sharp decrease occurs when the 
volume fraction of solid rock is equal to the volume fraction of melt, at which point the crystal/melt mixture behaves 
more like a liquid than a solid \citep{renner2001}.  For temperatures below the solidus, we assume that the viscosity 
of rock is controlled by the deformation rate due to volume diffusion \citep{karato95,SM2000}.  For ice I, we assume 
$\rho_{\rm iw}=1000$ kg/m$^3$ and that the viscous creep is accommodated by volume diffusion 
\citep{GoldsbyKohlstedt2001}.  Above the melting point, values for the viscosity of liquid water are used, and the 
rigidity is set to zero.  For the high pressure ice polymorphs, we assume $\rho_{\rm hpp}=1300$ kg/m$^3$, which 
accounts for the possible presence of ice phases II through VII \citep{Hobbs74}.  We assume that the viscosity of the 
hpp ices follows an approximate Newtonian flow law applicable to ices VI and VII, which are volumetrically dominant in 
the TRAPPIST-1 bodies \citep{barr18}.  For the shear modulus for hpp ices, we use an average value over the range for 
which experimental data exists \citep{shimizu96}.  For the density of iron, $\rho_{\rm Fe}=12,000$ kg/m$^3$, consistent 
with the density of Earth's inner core \citep{dziewonski81, unterborn18}.

{We compute the total amount of energy produced by tidal dissipation in each planet assuming once again that the 
planets are in synchronous rotation, so that the multiple tidal forcing frequencies are all equal to the orbital 
frequency. For this configuration the tidal dissipation is} \citep{segatz88},
\begin{equation}
   \label{viscel}
       \dot{E}_\mathrm{tidal} = - \frac {21} {2} {\rm Im}(k_2^*) \frac {R_{\rm pl}^5 n^5 e^2} {G} \, ,
\end{equation}
where Im($k_2^*$) is the imaginary part of $k_2^*$ and $G$ is the gravitational constant and $n$ is the orbital 
frequency (which is equal to the rotation frequency). Combining equations (\ref{eq:k2}), (\ref{eq:M1}), and 
(\ref{eq:M2}) to find the real and imaginary parts of $k_2^*$, we get
\begin{equation}
{\rm Re}(k_2^*) = \frac{3}{2}\frac{2 \beta (2 \beta + 19 M_1)}{(2 \beta + 19 M_1)^2 + (19 M_2)^2},
\end{equation}
\begin{equation}
{\rm Im}(k_2^*)=-\frac{3}{2}\frac{2 \beta (19 M_2)}{(2 \beta + 19 M_1)^2 + (19 M_2)^2},
\end{equation}
where $\beta=\bar{\rho} g R_\mathrm{pl}$.  The magnitude of $k_2^*$ is
\begin{equation}
|k_2^*|=\frac{3 \beta}{[(2 \beta + 19 M_1)^2 + (19 M_2)^2]^{1/2}}.
\end{equation}
The tidal quality factor is then computed as $Q={\rm Im}(k_2^*)/|k_2^*|$ \citep{segatz88}. Following the method of 
\citet{MignardI}, it is common to use the quantity $k_2 \Delta t$ to calculate tidal torques (e.g. \cite{bolmont15}), 
where $k_2$ is the amplitude of $k_2^*$ and $\Delta t$ is the time lag between the application of the tidal force and 
the planet's response to the forcing \citep{murray99}. The conversion from a $Q$-model to a time lag model is given by
\begin{equation}
k_2 \Delta t = \frac{|k_2^*|}{\nu Q}=\frac{{\rm Im}(k_2^*)}{\nu}.
\end{equation}
{In our simplified case $\nu=n$ and the computed value of $\Delta t$ is specifically tied to this forcing 
frequency. The relationship above should be used with extreme caution, because it is not applicable to cold, mostly 
solid planets because the forcing frequencies are generally very different from the natural frequencies inside the body 
\citep{Makarov2015}. As such, we provide the relationship for reference only.} We use the nominal planetary orbital and 
physical parameters together with the best-fit interior composition to  evaluate $\dot{E}_{\rm tidal}$, the tidal Love 
number and quality factor below.

\begin{figure*}
\centering
\includegraphics[width=40pc]{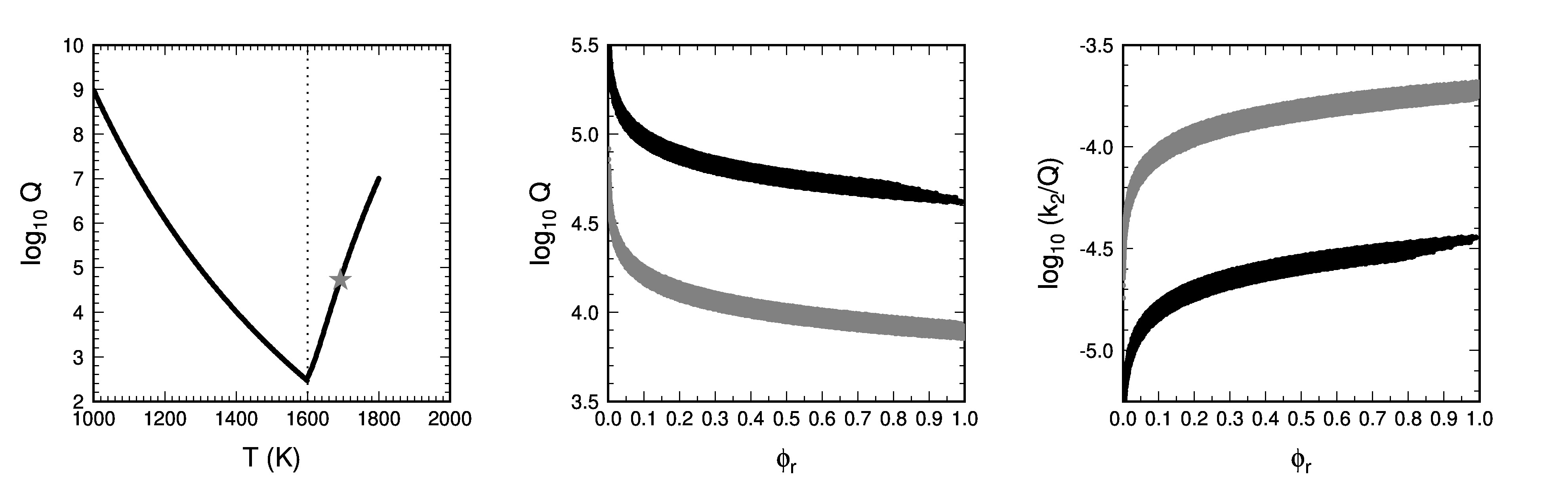}
\caption{(left) Tidal quality factor ($Q$) as a function of rock mantle temperature for a single representative model 
of TRAPPIST-1 b with $M_{\rm pl}=1.017 M_{\oplus}$, $R_{\rm pl}=1.121R_{\oplus}$, $\phi_{\rm r}=0.5$, $\phi_{\rm 
hpp}=0.41$, $\phi_{\rm iw}=0.01$, and $\phi_{\rm Fe}=0.08$. The value of $Q$ is minimized near the solidus temperature. 
 The gray star indicates $T=T_{\rm eq}$ for this model planet.  For temperatures lower than the solidus, the planet is 
too stiff to react to tidal flexing; conversely, for temperatures above the solidus, the presence of melt inhibits the 
dissipation of tidal energy as friction inside the planet. (middle) Values of $Q$ (on a log scale) for a range of 
models of TRAPPIST-1 b (black) and TRAPPIST-1 c (grey) as a function of $\phi_{\rm r}$. The $Q$ value is evaluated for 
$T=T_{\rm eq}$ for each model planet.  The width of the line represents uncertainties in $Q$ due to uncertainties in 
the mass and radius of each planet, as well as the relative volume fractions of rock, metal, and ice. (right) Same as 
the middle panel, but depicting $|k_2^*|/Q$ for each planet.}
\label{fig:QvsPhi}
\end{figure*}

We take into account all the possible interior structures of the TRAPPIST-1 planets as presented in the work of 
\citet{dobos19}. It varies from the lowest density case to the highest, according to the possible mass and radius pairs 
(taking into account their error bars) from Table~\ref{table:masses}. This covers even unlikely cases where one of the 
constituent materials are not present in the body (for example $\phi_\mathrm{r}=0$).

The left panel of Figure \ref{fig:QvsPhi} illustrates how the $Q$ value of TRAPPIST-1 b varies as a function of the 
rock mantle temperature.  Here, we have assumed the best-fit values for mass and radius: $M_{\rm b}=1.017M_{\oplus}$, 
$R_{\rm b}=1.121R_{\oplus}$.  We look at a single representative interior model with $\phi_{\rm r}=0.5$, $\phi_{\rm 
iw}=0.01$, $\phi_{\rm hpp}=0.41$, and $\phi_{\rm Fe}=0.08$, chosen because of its mixed ice/rock composition. The value 
of $Q$ changes by many orders of magnitude as the temperature of the rock mantle increases from 1000~K to 1800~K, so 
the behaviour of the interior of the planet in response to tides is worth discussing in detail. For an even more 
in-depth discussion see \cite{barr18}.

Tidal dissipation arises in a solid planet because the tidal forces drive the planet to deform, doing work against its 
own internal rigidity and self-gravity. For temperatures below the solidus, $\eta$ and $\mu$ are large, the planet is 
stiff and does not deform much in response to tidal forces. In this case, $Q$ can be enormously high, with values of 
$10^5$ or even $10^6$. At higher temperatures, the viscosity and rigidity of the bulk planet decrease, permitting more 
tidal deformation and dissipation.  However, as the planet's temperature increases above the solidus ($T_{\rm 
s}=$1600~K), the presence of melt in the planet's mantle causes the viscosity and rigidity to decrease.  Tidal 
dissipation becomes less efficient.  At temperatures well above the solidus, the viscosity and rigidity of the rock 
mantle are extremely low, so the internal rigidity of the planet is low, and tidal dissipation is small, leading once 
again to high $Q$ values. Tidal dissipation is maximized when the mantle is at the solidus temperature, and $Q\sim 
300$.  However, the equilibrium mantle temperature, at which the tidal heat flux is equal to the convective heat flux, 
is larger than $T_{\rm s}$; in the case of TRAPPIST-1 b this temperature is $T_{\rm eq}=1692$~K, which contains a melt 
fraction at which there is both strong tidal heating and efficient convective heat transport. For this structure and 
composition for planet b, $Q=5.65 \times 10^4$; {we also compute $|k_2^*| = 1.498$ and the Maxwell time $\tau_M = 
20.5$~d. The very high $|k_2^*|$ value is caused by the presence of partial melt in the mantle.}

To compare the behaviours of planets b and c, consider the $Q$ value of TRAPPIST-1 c calculated for a reference case 
with $M_{\rm pl}=1.156 M_{\oplus}$, $R_{\rm pl}=1.095R_{\oplus}$, $\phi_{\rm iw}=0.009$, $\phi_{\rm hpp}=0.328$, 
$\phi_{\rm r}=0.5$, $\phi_{\rm Fe}=0.163$. In this case, $T_{\rm eq}=1664$~K, cooler than planet b, and $Q=1.01\times 
10^4$, a factor of 5 lower than the quality factor for planet b; {for this planet $|k_2^*|=1.49$ and $\tau_M = 
26.9$~d.}

The middle panel of Figure \ref{fig:QvsPhi} illustrates how the $Q$ values for TRAPPIST-1 b and c vary as a function of 
the rock volume fraction for all possible interior states.  The width of the lines illustrate the uncertainty in $Q$ 
due to uncertainties in the masses and radii of the planets, as well as the amounts of rock/metal/ice in their 
interiors.  We find that TRAPPIST-1 b has $Q=(4-20)\times 10^4$ for cases where $\phi_\mathrm{r}>0.1$ (i.e., excluding 
the unlikely cases where the planet contains very little rock).  TRAPPIST-1 c has $Q = (0.7-3.7)\times 10^4$. The 
values 
of $k_2/Q$ range from $(0.075-0.37) \times 10^{-4}$ for TRAPPIST-1 b and $(0.4-2)\times 10^{-4}$ for TRAPPIST-1 c. This 
computed value of $k_2/Q$ for planet b is lower than what was predicted from the dynamical simulations in the previous 
sections, but the agreement between these two approaches is still reasonable. The level of disagreement between the two 
computed values of $k_2/Q$ for planet c is similar. {The corresponding time delays are then $k_2\Delta t=0.04$ sec 
to $0.75$ sec for planet b and $k_2\Delta t=0.6$ sec to $7$ sec for planet c.}

\section{Moments of inertia and precession frequencies of spin poles}
The interior models presented in the previous section can be used to compute the polar moment of inertia of the planets 
and compare their internal mass distribution with known values for solar system bodies. The combination of the moment 
of inertia and tidal parameters allows us to constrain the precession constants of the spin poles of these planets, 
which could have implications for climate cycles.

\begin{figure*}
\centering
\centerline{\includegraphics[width=30pc]{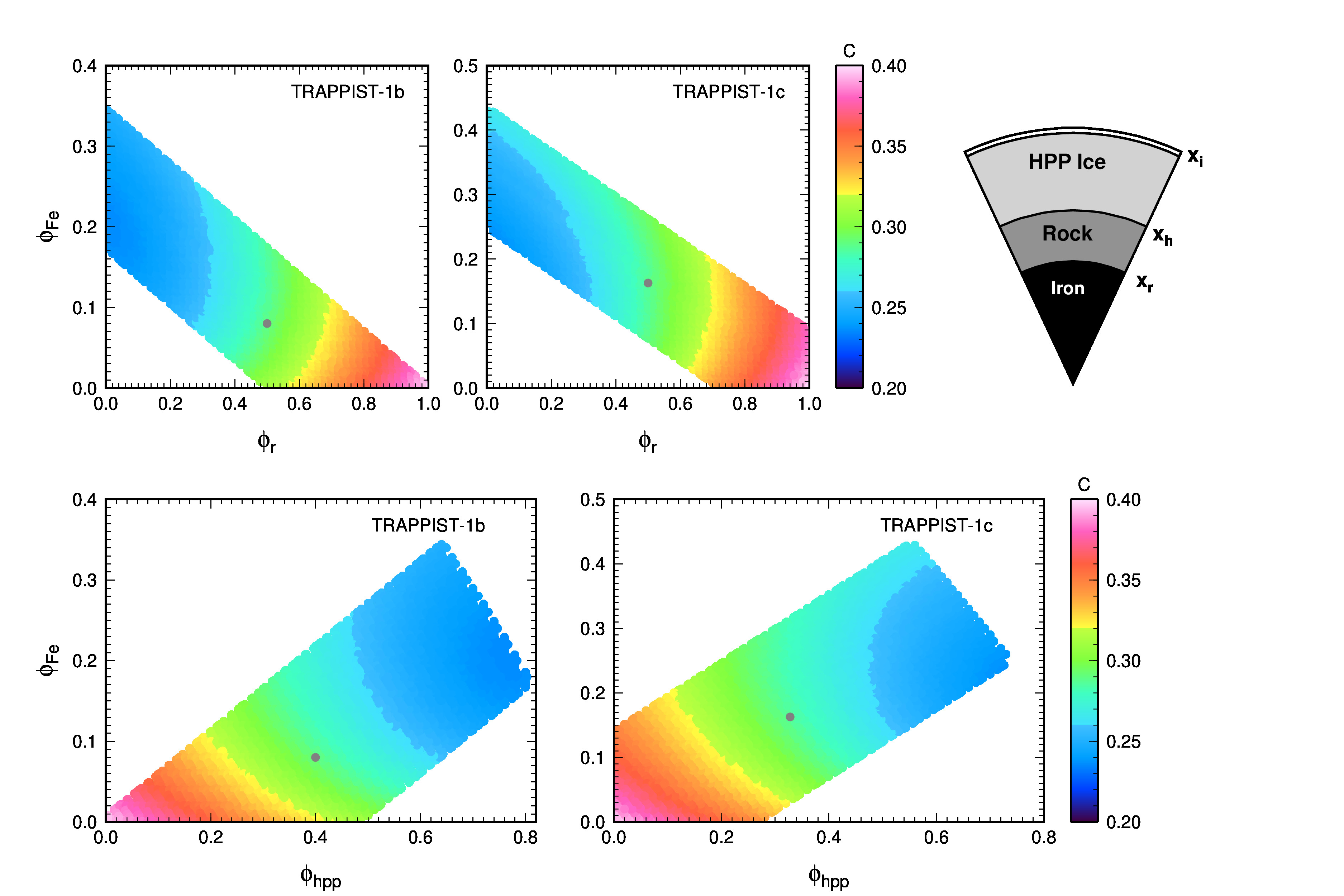}}
\caption{(top left) Colours show the values of the moment of inertia ($C$) for each valid structure for TRAPPIST-1 b as 
a function of $\phi_{\rm Fe}$ and $\phi_{\rm r}$.  Values range from $C \sim 0.25$ for rock- and iron-rich planets to 
$C\sim 0.4$ for pure rock planets.  Gray circle indicates the representative composition for TRAPPIST-1 b used to 
calculate $Q$ in the left panel of Figure \ref{fig:QvsPhi}.  (top middle) Same as top left panel, but for TRAPPIST-1 c. 
Gray circle indicates the representative compositional model for planet c. (top, right) Schematic illustration of the 
definitions of $x_r$, $x_h$, and $x_i$.  (bottom left) Values of $C$ for TRAPPIST-1 b as a function of $\phi_{\rm Fe}$ 
and $\phi_{\rm hpp}$.  (bottom right) Same as bottom left panel, but for planet c.
\label{fig:moi}}
\end{figure*}

The moment of inertia for a spherical planet about its principal rotation axis is \citep{turcotte02},
\begin{equation}
C=\frac{1}{M_{\rm pl}R_{\rm pl}^2}\int_{0}^{M_{\rm pl}} r^2 dm.
\end{equation}
For a uniform-density sphere, $C=2/5$, and values less than 2/5 indicate the concentration of high-density materials 
toward the centre of the planet.  Assuming the planets are fully differentiated, which seems likely given their high 
rock fractions \citep{friedson83} and tidal heat budgets \citep{barr18}, the moment of inertia integral can be 
evaluated for each distinct compositional layer, giving
\begin{equation}
C=\frac{2}{5} \bigg[z_\mathrm{Fe} x_\mathrm{r}^5 + z_\mathrm{r} (x_\mathrm{h}^5 - x_\mathrm{r}^5) + 
z_\mathrm{h}(x_\mathrm{i}^5-x_\mathrm{h}^5)+z_\mathrm{i} (1-x_\mathrm{i}^5)\bigg],
\end{equation}
where $z_\mathrm{Fe}=\rho_\mathrm{Fe}/\bar{\rho}$, $z_\mathrm{r}=\rho_\mathrm{r}/\bar{\rho}$, 
$z_\mathrm{h}=\rho_\mathrm{hpp}/\bar{\rho}$, $z_\mathrm{i}=\rho_\mathrm{i}/\bar{\rho}$, 
$x_\mathrm{r}=\phi_\mathrm{Fe}^{1/3}$, $x_\mathrm{h}= (\phi_\mathrm{Fe}+\phi_\mathrm{r})^{1/3}$, and 
$x_\mathrm{i}=1-x_\mathrm{r}-x_\mathrm{h}$ (see the top right panel of Figure~\ref{fig:moi} for an illustration of the 
definition of the $x$ values).

Figure \ref{fig:moi} illustrates how the moment of inertia vary for each valid interior structure for planets b and c. 
We find that $C$ ranges from $\sim 0.235$ (extremely centrally condensed) to 0.4 (a homogeneous sphere). For 
comparison, in our Solar System, the solid planetary bodies have $C \sim 0.31$ (Jupiter's moon Ganymede) to $0.38$ 
\citep[Jupiter's moon Callisto;][]{schubert04}; Earth has $C=0.33$ \citep{turcotte02} and Mars has $C=0.366$ 
\citep{BertkaFeiEPSL} while the Moon has $C = 0.395$ \citep{Bolt1960}. The gas giants have $C \sim 0.25$ \citep{Hel14}. 

It should be noted that simply assuming a terrestrial value for the moment of inertia for all of the planets can cause 
the tidal torque to be overestimated by $\sim$20 to 50\% \citep{MignardI}.  From our compositional models for the 
TRAPPIST-1 planets, it seems that $C$ can be as low as 0.235, which can arise for planets whose interiors are composed 
of ice and iron and have very little rock.  Taking into account geochemical arguments about the ratios of rock to metal 
\citep{unterborn18} we may be able to rule out such extreme cases. For the representative cases (gray circles in Figure 
\ref{fig:moi}) we get $C=0.289$ for planet b and $C=0.286$ for planet c.

The moment of inertia values of planets b and c can be used to compute the precession frequencies of their spin poles. 
Tidal evolution should have placed the planets' obliquities in Cassini state 1, which has an obliquity near 0$^\circ$ 
\citep{Col66}, and their rotations into the synchronous state. Then, due to rotational and tidal deformation, the 
expected value of $J_2$ is \citep{CR13}

\begin{equation}
    J_2 = k_f \frac{n^2 R_{\rm pl}^3}{GM_{\rm pl}} + \frac{k_2}{2}\frac{M_*}{M_{\rm pl}}\Bigl(\frac{R_{\rm 
pl}}{a}\Bigr)^3,
\end{equation}
where we have assumed $e \approx 0$ and a zero obliquity. Our choice for the eccentricity is justified because for all 
the planets $e<0.01$ and the leading term in eccentricity for $J_2$ is $1+\frac{3}{2}e^2$. In the expression for $J_2$, 
$n$ is the mean motion, and $k_f$ is the fluid Love number, which is related to the moment of inertia via the 
Darwin-Radau relationship,

\begin{equation}
    C = \frac{2}{3}\Bigl(1-\frac{2}{5}\sqrt{\frac{4-k_f}{1+k_f}}\Bigr).
\end{equation}
The precession frequency of the spin pole is given by \citep{Col66}

\begin{equation}
    \alpha = \frac{3}{2}n\frac{J_2}{C}(1-e^2)^{-3/2},
\end{equation}
where we assumed synchronous rotation.

\begin{figure}
\centerline{\includegraphics[width=\columnwidth]{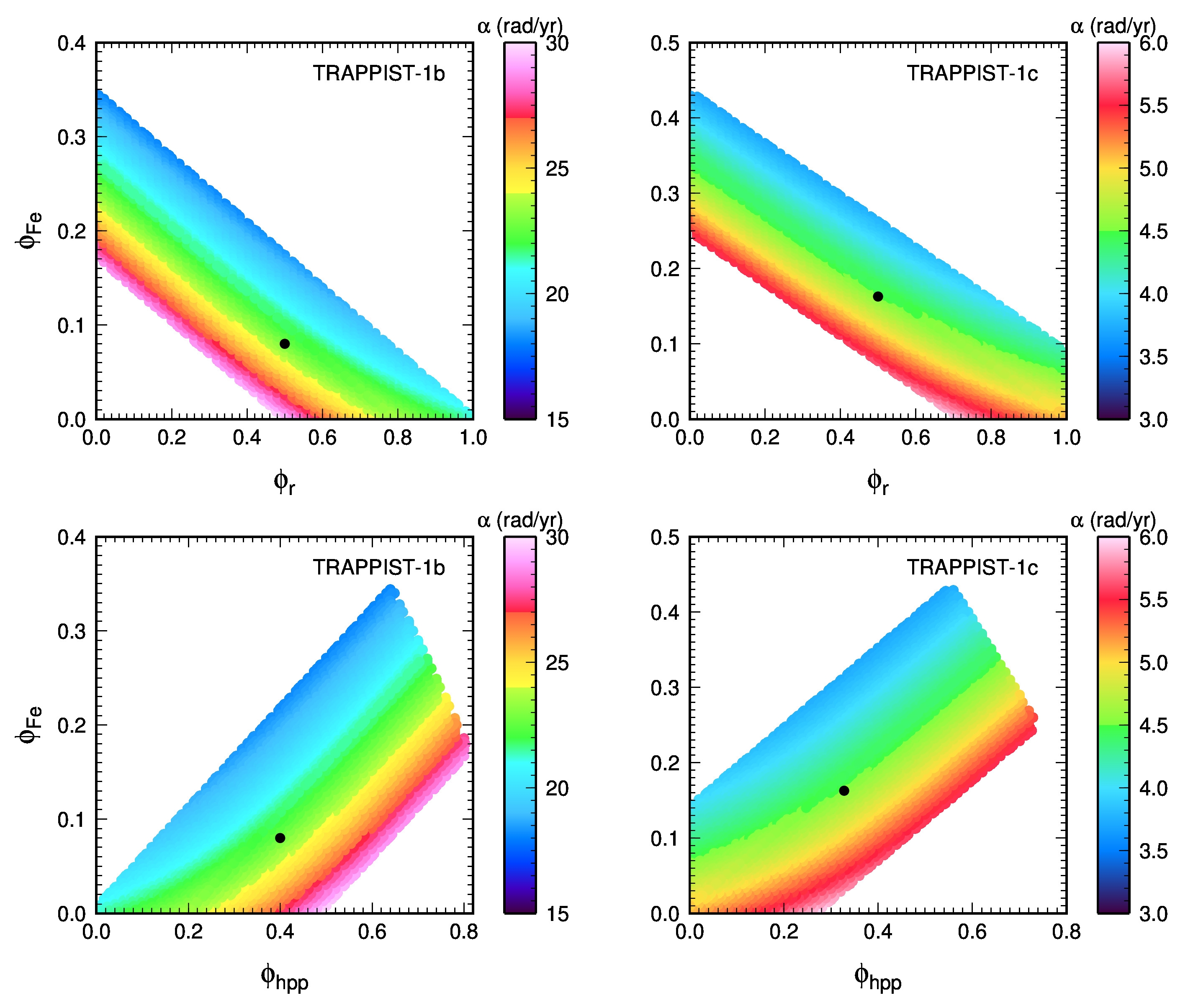}}
\caption{(top left) Colours show the values of the precession frequency for the spin pole for each valid structure for 
TRAPPIST-1b as a function of $\phi_\mathrm{Fe}$ and $\phi_\mathrm{r}$. Black circle indicates the representative 
composition for TRAPPIST-1 b used to calculate $Q$ in the left panel of Figure \ref{fig:QvsPhi}.  (top right) Same as 
top left panel, but for TRAPPIST-1 c.  (bottom left) Values of $\alpha$ for TRAPPIST-1 b as a function of 
$\phi_\mathrm{Fe}$ and $\phi_\mathrm{hpp}$.  (bottom right) Same as bottom left panel, but for planet c.}
\label{fig:precession}
\end{figure}

Adopting the representative composition for TRAPPIST1-b, we obtain $J_2 = 2.92 \times 10^{-3}$ and that $\alpha = 23$ 
rad/yr, while for the representative case for planet c we have $J_2 = 9.6\times 10^{-4}$ and $\alpha = 4.77$ rad/yr. 
The spin precession periods for both planets are shorter than an Earth year. These are faster than the eccentricity 
precession frequencies which exclude chaotic obliquity variations. For a full range of possible $\alpha$ values 
obtained for the different possible interior structures, see Figure \ref{fig:precession}. For planet b, $\alpha$ varies 
between 16 and 31 rad/yr, but for planet c the range is much smaller: $3.5<\alpha<6.5$ rad/yr. The precession 
frequencies of the ascending nodes of these planets are expected to be similar in magnitude, but in the opposite 
direction, to their apsidal precession frequencies i.e. approximately $-29000$ ''/yr. As such, the spin precession 
frequencies of the poles are much faster than their nodes, and, apart from obliquities very close to 90$^\circ$, we do 
not expect these planets to be in a secular spin-orbit resonance or to have a chaotic obliquity evolution.

\section{Discussion and Conclusions}
Although the TRAPPIST-1 planets experience tidal heat fluxes similar in magnitude to tidal heat fluxes experienced by 
Solar System objects like Io and Enceladus \citep{spencer06, howett11, barr18, dobos19}, the values of tidal quality 
factor, $Q$, are much larger than values estimated for Solar System objects.  This is because the conversion of tidal 
energy to thermal energy is inefficient in a partially molten planetary mantle: the rigidity of the planet has 
decreased, so the planet experiences less internal friction.  In the absence of prior estimates of $Q$ for the 
TRAPPIST-1 planets, $N$-body simulations evaluating the stability of the orbits in the TRAPPIST-1 system have assumed 
Earth-like values: $Q \sim$ a few hundred \citep{gillon17, bolmont15}.  This may lead to inaccurate estimates of the 
dynamical lifetime of the system \citep{gillon17, tamayo17}.  

Here we have combined long-term N-body simulations with sophisticated tidal and thermal modelling. Specifically, for 
the N-body simulations we used initial conditions that are predicted from the formation of the TRAPPIST-1 system to 
study the system's dynamical stability. We found that the median instability time is close to 30~Myr, consistent with 
predictions of the dynamical lifetime of other such multi-resonant systems \citep{I17}. Tidal secular theory predicts 
that the majority of the tidal dissipation of the system occurs in the innermost two planets. We used this theory and 
the dynamical instability timescale to constrain the tidal parameters of the inner two planets. We have found lower 
limits for the tidal parameter $k_2/Q$, having values above $2 \times 10^{-4}$  for TRAPPIST-1 b and $10^{-3}$ for 
TRAPPIST-1 c.

Additionally, we applied multi-layered viscoelastic deformation models of the type used to estimate tidal dissipation 
and stresses in the Earth and the Galilean and Saturnian satellites \citep{sabadini82, segatz88, roberts08, wahr09}, to 
provide more realistic estimates of Im($k_2)$ and $Q$ for the TRAPPIST-1 planets. We found that the fiducial interior 
model gives lower values for the $k_2/Q$ tidal parameter of planets b and c (in the range of $(0.08 - 0.37) \times 
10^{-4}$ and  $(0.4 - 2) \times 10^{-4}$, respectively) than obtained from the dynamical models.

The agreement is better if we increase the maximum eccentricity damping timescale from 30~Myrs to 100~Myrs because this 
would decrease the allowed minimum value of $k_2/Q$ and for it to be more consistent with those calculated from the 
interior models. 

{In this work, however, we made several simplifying assumptions, first and foremost that all the planets are 
assumed to be synchronous throughout most of their evolution. {As the planets spin down, they are expected to be 
temporarily trapped in spin-orbit resonances. From our numerical simulations the eccentricities of the TRAPPIST-1 
planets remain at values near 0.01. As a result, planet c almost certainly spun down to the synchronous state, the 
probabilities of its capture in the 2:1 and 3:2 spin states being 0 and 0.02, correspondingly. At the 
same time, planet b could have been captured in the 2:1 spin-orbit resonance with a probability of as high as 0.2; 
its capture in the 3:2 spin-orbit resonance was certain had it skipped past the 2:1 \citep{makarov18}. It ensues from 
equation (156) in \citet{be2019} that, while in the synchronous and 3:2 spin-orbit resonances $\dot{e}<0$ always, in 
the 2:1 spin-orbit resonance this rate is positive (up to some caveats). Thus, had planet b been trapped in the 
2:1 spin-orbit resonance, this planet’s eccentricity was not decreasing but increasing during that capture.
Should such eccentricity pumping} have occurred it begs the question if this could have affected the 
stability of the whole system during the initial higher-eccentricity phase. Such an eccentricity increase would also 
increase the AMD of the system and this increase would need to be compensated by stronger damping in the other planets 
because the system did not undergo a global instability caused by excess AMD. We reserve an investigation of 
eccentricity damping versus pumping for future work.} 

{The temporary trapping in the 3:2 spin-orbit resonance state has another consequence. In these higher spin-orbit 
resonance states the tidal dissipation rate is orders of magnitude higher than in the synchronous state, so that 
captured planets gradually heat up further, potentially alter their rheology, subsequently pop out of these spin-orbit 
resonant states and continue to despin towards the synchronous state. However, a heating of the interior appears to 
{\it decrease} the efficiency of tidal eccentricity damping, leading to a paradox as to how the system damped the 
excess eccentricity at all. It is possible that most of the tidal damping happened in the 3:2 spin-orbit resonance 
configuration which could have changed the tidal parameters computed both from the dynamics and interior models, and 
when taking this into account the agreement between the two could be better.} 

Despite the somewhat different outcomes from the two models, we encourage the application of this 
two-method approach to constrain the tidal parameters of planets in other multi-resonant systems.

\section*{Acknowledgements}
{The authors thank the reviewer Michael Efroimsky for useful questions and constructive comments that improved 
the quality of this manuscript.} R.~B. acknowledges financial assistance from the Japan Society for the Promotion of 
Science (JSPS) International Joint Research Fund (JP17KK0089) and JSPS Shingakujutsu Kobo (JP19H05071). V.~D. is 
supported by the Hungarian National Research, Development, and Innovation Office (NKFIH) grants K-119993, K-115709, and 
GINOP-2.3.2-15-2016-00003. A.~C.~B. acknowledges support from NASA Habitable Worlds 80NSSC18K0136.

\bibliographystyle{mnras}
\bibliography{ref}
\end{document}